\documentclass[showpacs,aps,prl,twocolumn,superscriptaddress]{revtex4}

\usepackage{graphicx}

\begin{document}


\title{Optical probe of carrier doping by X-ray irradiation in organic dimer Mott insulator $\kappa$-(BEDT-TTF)$_{2}$Cu[N(CN)$_{2}]$Cl}


\author{T. Sasaki}
\email{takahiko@imr.tohoku.ac.jp}
\author{N. Yoneyama}
\author{Y. Nakamura}
\author{N. Kobayashi}
\affiliation{Institute for Materials Research, Tohoku University, Sendai 980-8577, Japan}
\author{Y. Ikemoto}
\author{T. Moriwaki}
\author{H. Kimura}
\affiliation{SPring-8, Japan Synchrotron Radiation Research Institute, Sayo, Hyogo 679-5198, Japan}


\date{\today}

\begin{abstract}
We investigated the infrared optical spectra of an organic dimer Mott insulator $\kappa$-(BEDT-TTF)$_{2}$Cu[N(CN)$_{2}$]Cl, which was irradiated with X-rays.
We observed that the irradiation caused a large spectral weight transfer from the mid-infrared region, where interband transitions in the dimer and  Mott-Hubbard bands take place, to a Drude part in a low-energy region; this caused the Mott gap to collapse. 
The increase of the Drude part indicates a carrier doping into the Mott insulator due to irradiation defects.
The strong redistribution of the spectral weight demonstrates that the organic Mott insulator is very close to the phase border of the bandwidth-controlled Mott transition.
\end{abstract}

\pacs{71.30.+h, 71.27.+a, 78.30.Jw}

\maketitle


Metal-insulator (MI) transitions are of considerable importance in the strongly correlated electron systems.
Among the various types of MI transitions, the Mott transition due to electron-electron interactions is the most attractive phenomenon \cite{Imada}.
A Mott insulator derives from the large on-site Coulomb energy with respect to the bandwidth. 
There are two types of Mott transitions.
In the first type, the strength of the interaction changes and the carrier filling is maintained at a commensurate value (bandwidth-controlled Mott transition), while in the second type, the carrier is introduced to obtain the required density and the strength of the interaction is constant (filling-controlled Mott transition).  
In molecular conductors, the former type of transition occurs, while the latter type occurs typically in transition metal oxides.  

Organic charge-transfer salts based on a donor molecule bis(ethylenedithio)-tetrathiafulvalene (abbreviated BEDT-TTF) have been recognized as one of highly correlated electron systems \cite{Powell}.  
Among them, $\kappa$-(BEDT-TTF)$_{2}$$X$ with $X =$ Cu(NCS)$_{2}$, Cu[N(CN)$_{2}$]$Y$ ($Y =$ Br and Cl), etc. has attracted considerable attention as a bandwidth-controlled Mott transition system because of its strong dimer structure consisting of two donor molecules, which effectively makes the conduction band a half-filling band \cite{Powell,Kanoda,Kino}.
Recently, a study was conducted on the effect of X-ray-irradiation-induced carrier doping on the dc conductivity of a Mott insulator $\kappa$-(BEDT-TTF)$_{2}$Cu[N(CN)$_{2}$]Cl (hereafter $\kappa$-Cl) \cite{Sasaki1}.  
It was observed that there was a large decrease in resistivity due to irradiation at room temperature.  
Moreover, the temperature variation of the resistivity down to 50 K showed a metal-like temperature dependence.  
The irradiation-induced molecular defects \cite{Zuppirol,Analytis} were expected to cause local imbalances in the charge transfer in the crystal and the effective carrier doping of the half-filled dimer Mott insulator.

In this letter, we report the effect of X-ray irradiation on the infrared optical spectra of $\kappa$-Cl in order to understand the effect of carrier doping, as proposed in a previous dc transport investigations \cite{Sasaki1}.


Single crystals of $\kappa$-Cl were grown by a standard electrochemical oxidation method.  
Five samples (No. 1 to No. 5) were irradiated individually at 300 K using a nonfiltered tungsten target at 40 kV and 20 mA.  
The dose rates in the present experimental conditions were approximately 0.5 MGy/hour and this value was determined on the basis of comparisons with a previous report \cite{Analytis,dose}.
The irradiation time was the sum of multiple exposures at 300 K.
The polarized reflectance spectra in the mid-infrared (IR) region (600 - 8000 cm$^{-1}$) were measured along the electric field direction $E$ parallel to the $a$-axis ($E{\|}a$) and the $c$-axis ($E{\|}c$) with a Fourier transform microscope-spectrometer.  
Synchrotron radiation light at BL43IR in SPring-8 was used for measurements in the far-IR region (100 - 600 cm$^{-1}$) \cite{Ikemoto}.  
The reflectivity was determined by comparison with a thin gold film evaporated partly on the crystal surface.
The optical conductivity was calculated by a Kramers-Kronig analysis of the reflectivity.


\begin{figure}
\includegraphics[viewport=2cm 5.5cm 17cm 24cm,clip,width=0.9\linewidth]{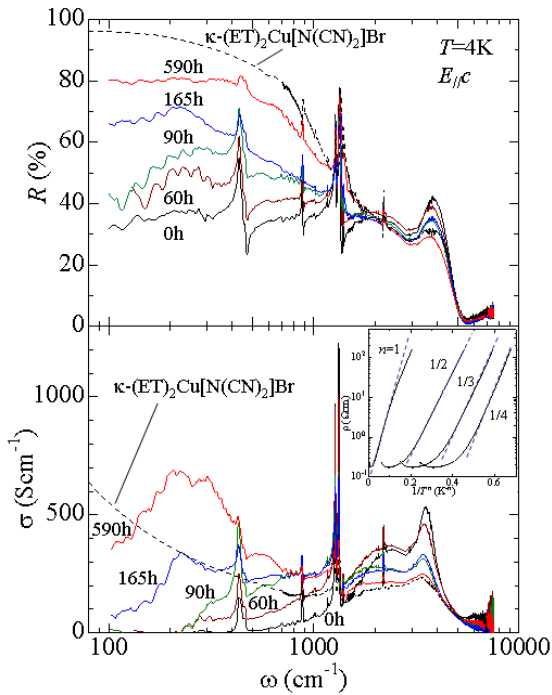}
\caption{(color online) Optical reflectivity (upper part) and conductivity (lower part) spectra of $\kappa$-Cl before and after X-ray irradiation. The dashed curves are the results of the nonirradiated $\kappa$-Br. The inset in the lower part shows the logarithmic plot of dc resistivity as a function of $1/T^{n}$, where $n =$ 1, 1/2, 1/3, and 1/4, after 325 h irradiation .}
\end{figure}

Figure 1 shows the reflectivity and conductivity spectra at 4 K in $E \|{c}$ of $\kappa$-Cl before and after X-ray irradiation.  
The spectra are presented for the samples No. 1 (0 hour), No. 2 (60 and 90 h), No. 3 (165 h), and No.5 (590 h). 
The spectra of nonirradiated $\kappa$-(BEDT-TTF)$_{2}$Cu[N(CN)$_{2}$]Br (hereafter $\kappa$-Br), which is a superconductor with $T_{\rm c}$ of 11 K, are shown for the purpose of comparison. 
Both, the spectra of nonirradiated $\kappa$-Cl and $\kappa$-Br, were qualitatively in good agreement with results of the previous reports \cite{Kornelsen,Eldridge,Sasaki2}.
Large absorption peaks at 2250 cm$^{-1}$ and 3350 cm$^{-1}$ in $\kappa$-Cl were attributed to interband transitions in the dimer bands and Mott-Hubbard bands, respectively \cite{Faltermeier}.
In addition, an optical gap corresponding to a Mott-Hubbard gap appeared approximately below 1000 cm$^{-1}$ \cite{Kornelsen}.  
In the case of the superconductor $\kappa$-Br, interband transitions were weak and a Drude response appeared in the far-IR region.  

The magnitude of the interband transitions of $\kappa$-Cl is reduced by X-ray irradiation.  
The conductivity of the broad absorption peaks decreases with increasing the irradiation time. 
It should be noted that the irradiation did not induce the change of the absorption peak frequencies which correspond to the interband transition energies. 
This means that the fundamental electronic parameters, i.e, the intra-dimer transfer energy and the effective Coulomb energy of the dimer site are not affected by X-ray irradiation.
The reduction in the spectral weight (SW) in the mid-IR region due to irradiation was compensated by the shift of the SW to the far-IR region.  
Thus, the SW of the Mott-Hubbard gap increased with the irradiation time.  
However, we did not observe a Drude absorption peak at $\omega =$ 0 when the irradiation time was long.  
A large decrease in the dc resistivity and metallic conduction properties were observed after X-ray irradiation \cite{Sasaki1}, however, the resistivity showed a weak insulating behavior at low temperatures, as indicated in the inset in the lower part of Fig. 1.
Such weak insulating behavior is consistent with disappearance of the Drude peak.
An interesting problem is the determination of the nature of the carriers responsible for large SW transfers.  
Possible origin of no Drude peak is the incoherent motion of carriers caused by the electron correlation effect \cite{Arima}, which has been discussed as a so-called bad metal in the strongly correlated materials including molecular conductors at high temperatures \cite{Powell,Sasaki2,Takenaka}.  
We also need to consider the randomness effect induced by X-ray irradiation. 
Inhomogeneity often leads to electron localization.  
Electron localizations may shift the SW of the Drude peak to regions with high energy \cite{Dressel}. 
In fact, the increase of dc resistivity at low temperatures follows not activation-type behavior but localization type behavior.  
The inset of Fig. 1 shows the dc resistivity of $\kappa$-Cl after 325 h of X-ray irradiation.  
The dc resistivity is plotted logarithmically as a function of $1/T^{n}$ with $n =$ 1, 1/2, 1/3, and 1/4. 
The linear dependence in this plot can be stated as $\rho(T) \propto \exp[(T_{0}/T)^{n}]$.  
The plot with $n =$ 1/2 shows better linear dependence than those with $n =$ 1 for a thermal activation-type, 1/3 for variable range hopping (VRH) in two dimensions, and 1/4 for VRH in three dimensions.
The case where $n = 1/2$ has been known to appear in localized states that have electron-electron interactions.  
This being the case leads to the formation of a Coulomb gap \cite{Efros}.  
Both the dc resistivity and far-IR conductivity suggest that the weakly disordered metal state with a Coulomb gap, in which electron-electron interactions occur, is induced from the Mott insulating state by X-ray irradiation.  
Thus, it is difficult to define clearly the SW of the far-IR region.  
Therefore, we will not discuss the distinction between the incoherent state of the carriers and the randomness effect any further.  
We define a Drude part in the far-IR region as the SW that is transferred across the isosbestic point ($\sim$ 1700 cm$^{-1}$) from the mid-IR region where the interband transition appears.

\begin{figure}
\includegraphics[viewport=4cm 10.5cm 17cm 21cm,clip,width=0.9\linewidth]{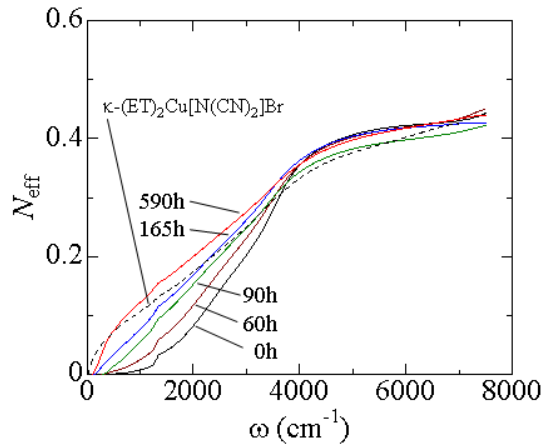}
\caption{(color online) Frequency dependence of the effective number of carriers $N_{\rm eff}$ of $\kappa$-Cl at 4 K before and after irradiation. The dashed curve shows the results of nonirradiated $\kappa$-Br.}
\end{figure}

In order to quantitatively evaluate the SW transfer by X-ray irradiation, we define the effective number of carriers as 
\begin{equation}
N_{\rm eff}(\omega) = \frac{2m_{\rm 0}}{\pi e^{2} N} \int_{0}^{\omega} \sigma (\omega')d\omega', 
\end{equation}
where $m_{\rm 0}$ is the free electron mass and $N$ is the number of BEDT-TTF dimers per unit volume.  
Figure 2 shows the effective carrier number $N_{\rm eff}(\omega)$ of X-ray-irradiated $\kappa$-Cl at 4 K. 
The $\omega$ dependence of $N_{\rm eff}$ shows how the SW is redistributed with increasing the irradiation time.  
In high $\omega$ region, $N_{\rm eff}(\omega)$ tends to saturation so as to approach the value of 0.45 -- 0.5.  
The saturation value, which represents a sum of each SW of the interband transitions, the Drude part, and molecular vibrations, seems to be nearly independent of the irradiation time. 
In other words, the sum of the SWs of these contributions is nearly conserved even with a variation in the irradiation time.  
This feature of the conservation is also valid for the superconductor $\kappa$-Br.  
The saturation value of $N_{\rm eff}(\omega)$ suggests that the optical band mass $m_{b, {\rm opt}}$ is approximately (2.0 -- 2.2)$m_{\rm 0}$, which is evaluated using $m_{b, {\rm opt}}$ instead of $m_{\rm 0}$ based on the assumption that the saturated value of $N_{\rm eff}(\omega)$ should be unity.  
The obtained value of $m_{b, {\rm opt}}$ is in agreement with its calculated (2.3 $m_{\rm 0}$ \cite{Merino}) and experimental (2.5 $m_{\rm 0}$ \cite{Faltermeier}) values in previous reports.  

\begin{figure}
\includegraphics[viewport=2.5cm 10.5cm 16.5cm 20.5cm,clip,width=1.0\linewidth]{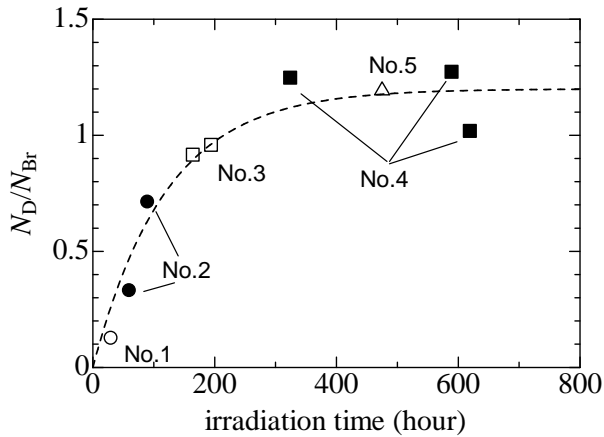}
\caption{Irradiation time dependence of the effective number $N_{\rm D}$ of charge carries in the Drude part.  The value is plotted after normalization by $N_{\rm Br}$ for the superconductor $\kappa$-Br. The dashed curve is an exponential function fitted to the data points. For the details, see the text.}
\end{figure}

The number of carriers in the Drude part is defined as
\begin{equation}
N_{\rm D} = N_{\rm eff}(\omega_{\rm iso})_{t_{\rm irr}} - N_{\rm eff}(\omega_{\rm iso})_{t_{\rm irr} = 0}, 
\end{equation}
where $\omega_{\rm iso}$ is the frequency (1700 cm$^{-1}$) at the isosbestic point, and $t_{\rm irr}$ is the irradiation time.  
Figure 3 shows the dependence of the effective number $N_{\rm D}$ of carriers in the Drude part on irradiation time.  
In the plot, $N_{\rm D}$ is normalized by $N_{\rm Br}$ which is obtained by replacing $N_{\rm eff}(\omega_{\rm iso})_{t_{\rm irr}}$ in Eq. (2) with the results of $\kappa$-Br.  
The observed irradiation time dependence can be fitted roughly with a general exponential function for collision damping, $N_{\rm D}/N_{\rm Br} = (N_{\rm D}/N_{\rm Br})_{0}[1-\exp(-t_{\rm irr}/t_{0})]$, where $t_{0}$ is proportional to the inverse of a cross section and $(N_{\rm D}/N_{\rm Br})_{0}$ is the saturation value reached at a long $t_{\rm irr}$.
The dashed curve in Fig. 3 represents the result of fitting to the formula with $(N_{\rm D}/N_{\rm Br})_{0} =$ 1.2 and $t_{0} =$ 120 hours. 
Fairly good agreement between the data and the fitting also supports that the number of the carries in the Drude part is proportional to the concentration of the irradiation defects.
In the initial irradiation at $t_{\rm irr} <$ $\sim$ 100 hours, $N_{\rm D}$ increases almost linearly with the irradiation time as is similar to the rough approximation of the formula. 
Such a linear increase in $N_{\rm D}$ has been observed in transition metal oxides \cite{Arima,Katsufuji}, and it is consistent with the results of theoretical studies \cite{Dagotto,Lee}.
For a perovskite-type $R_{1-x}$Ca$_{x}$TiO$_{3}$ ($R =$ rare earth) Mott insulator, the slope of the linear dependence of $N_{\rm D}$ as a function of the hole doping by substituting Ca with $R$ increases critically as the system approaches the bandwidth-controlled Mott transition \cite{Katsufuji}.  
It is noted that the evolution value of $N_{\rm D}/N_{\rm Br}$ in the irradiated $\kappa$-Cl is large in comparison with the results of $R_{1-x}$Ca$_{x}$TiO$_{3}$.  
According to a model of the carrier doping by irradiation \cite{Sasaki1}, the number of induced carriers may be almost equal to the number of molecular defects. 
The actual defect concentration by X-ray irradiation is expected to be only a few percent of molecules \cite{Zuppirol}.
This suggests that the organic Mott insulator $\kappa$-Cl is located very closely in the Mott transition because large evolution of $N_{\rm D}$ is caused by small number of the carrier doping due to irradiation defects.
The variation in $N_{\rm D}(t_{\rm irr})$ when the bandwidth is being manipulated should be studied in the future in order to accurately compare the results of this study with the theoretical and experimental results of the inorganic systems.  
In addition, the number of carriers induced by irradiation should be determined quantitatively by Hall effect measurements.
 
\begin{figure}
\includegraphics[viewport=2.5cm 5.5cm 17.5cm 23cm,clip,width=0.9\linewidth]{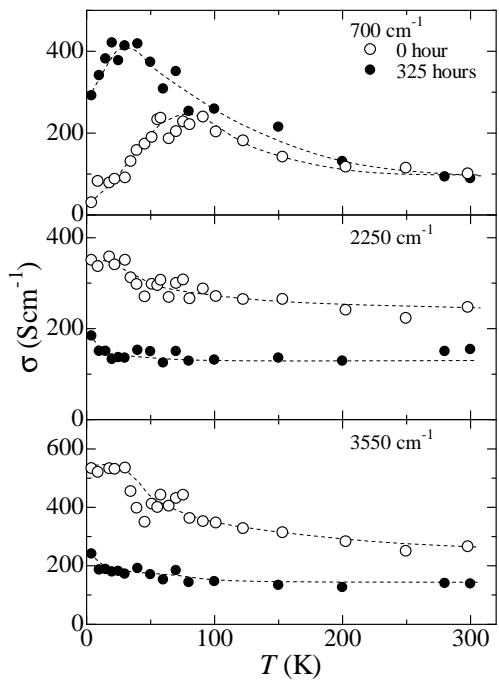}
\caption{Temperature dependence of the optical conductivity at 700, 2250, and 3550 cm$^{-1}$ for $E \|{c}$ of sample No. 4 after 325 h of irradiation.  The data at 0 h is obtained by a reanalysis of the previous results reported in Ref. \cite{Sasaki2}.  The dashed curves are provided as guides.}
\end{figure}

Figure 4 shows the temperature dependence of the optical conductivities at 700, 2250, and 3550 cm$^{-1}$ after 325 h of irradiation of sample No.4 and of the nonirradiated sample, as reported previously \cite{Sasaki2}.  
The magnitude at these frequencies reflects the redistribution of the SW in the Drude part (700 cm$^{-1}$) and the interband transitions in Mott-Hubbard bands (2250 cm$^{-1}$) and dimer bands (3550 cm$^{-1}$) \cite{Faltermeier}.  
In the case of the nonirradiated sample, the interband transitions evolve and the Drude part is suppressed below $T_{\rm ins} \simeq$ 50-70 K, which is the temperature at which the Mott gap starts to open.  
Antiferromagnetic (AF) ordering occurs at $T_{\rm N} \simeq$ 27 K and has been investigated by $^{1}$H-NMR and magnetic susceptibility measurements \cite{Miyagawa}.  
However, no indication of the AF ordering at $T_{\rm N}$ has appeared in the optical spectra and dc conductivity. 
After irradiation for 325 h, the magnitude of the interband transitions decreases and no reduction of the conductivities appears at $T_{\rm ins}$.  
A subtle change, however, can be seen in the conductivities around 25 K, indicating a small depression in the Drude part; this temperature is close to $T_{\rm N}$ of the nonirradiated sample.  
This indicates that AF correlations exist between the induced carriers.  
The coexistence of AF ordering and metallic conduction has been discussed theoretically for lightly doped cuprates \cite{Machida}.
Experimental investigations of the magnetic properties of the irradiated samples are necessary to know the nature of the doped correlated carriers.

In summary, we investigated the effect of X-ray irradiation on the infrared optical spectra of an organic dimer Mott insulator $\kappa$-Cl.
There is considerable SW transfer from the mid-IR region to a Drude part in a low-energy region followed by the collapse of the Mott gap; this indicates that irradiation can be used for effective doping of carriers. 
Finally, we would like to discuss the possible applications of carrier doping by irradiation. 
By using an X-ray microbeam, it is possible to fabricate electrical circuits and metallic dots for organic Mott insulators.  
Preliminary metallic pattern fabrication has been demonstrated \cite{Yoneyama}.

The authors thank T. Tohyama, S. Iwai, and K. Yakushi for valuable discussions.
The SR experiments were performed at SPring-8 with the approval of JASRI (2007B1150 and 2008A1121).
This work was partly supported by a Grant-in-Aid for Scientific Research (No. 16076201, 17340099, 18654056 and 20340085) from MEXT and JSPS, Japan.


\begin{thebibliography}{}
\bibitem{dose} We confirmed the dose rate on the basis of a comparison of the reduction of the superconducting transition temperature of $\kappa$-(BEDT-TTF)$_{2}$Cu(NCS)$_{2}$ irradiated under the present condition and the results reported in Ref. \cite{Analytis}.
\end{thebibliography}

\end{document}